\documentclass[twocolumn,showpacs,amsmath,amssymb,superscriptaddress]{revtex4}
\usepackage{amsmath,amssymb,color,latexsym,natbib,graphicx}

\usepackage{dcolumn}
\usepackage{bm}% bold math
\def\ADD#1{{\textcolor{black}{#1}}}    % comments, questions..
\newcommand{\p} {\partial}
\def\eg{{\it e.g.}\ } 
\def\ie{{\it i.e.}\ }

\def\vv{{\bf v}}
\def\D{{\bf d_K}}
\def\d_M{{\bf d_M}}
\def\fz{{\bf f_0}}
\def\rr{{\bf r}}
\def\zz{{\bf z}}
\def\bb{{\bf b}}
\def\xx{{\bf x}}
\def\vA{{\bf v_A}}
\def\Bz {{\bf B_0}}
\def\be{\begin{equation}}
\def\ee{\end{equation}}
\def\ba{\begin{eqnarray}}
\def\ea{\end{eqnarray}}

\def \pmbtext#1{\leavevmode
     \setbox0\hbox{#1}
     \kern0,4pt \copy0 \kern-\wd0
     \kern-0,2pt \raise0,3pt \box0 }
     
\begin{document}

\preprint{1}

\title{Exact Relation \ADD{with Two-point Correlation Functions and Phenomenological Approach} for Compressible Magnetohydrodynamic Turbulence}

\author{Supratik Banerjee}
\affiliation{Univ Paris-Sud, Institut d'Astrophysique Spatiale, UMR 8617, b\^at. 121, F-91405 Orsay, France}
\author{S\'ebastien Galtier}
\affiliation{Univ Paris-Sud, Institut d'Astrophysique Spatiale, UMR 8617, b\^at. 121, F-91405 Orsay, France}
\affiliation{Institut universitaire de France, 103, boulevard Saint-Michel, 75005 Paris, France}

\date{\today}
%%%%%%%%%%%%%%%
\begin{abstract}
Compressible isothermal magnetohydrodynamic turbulence is analyzed under the assumption of statistical homogeneity and in the asymptotic 
limit of large kinetic and magnetic Reynolds numbers. Following Kolmogorov we derive an exact relation for some two-point correlation 
functions which generalizes the expression recently found for hydrodynamics. We show that the magnetic field brings new source and flux terms 
into the dynamics which may act on the inertial range similarly as a source or a sink for the mean energy transfer rate. The introduction of a 
uniform magnetic field simplifies significantly the exact relation for which a simple phenomenology may be given. A prediction for axisymmetric 
energy spectra is eventually proposed. 
\end{abstract}
%%%%%%%%%%%%%%%
\maketitle

%%%%%%%%%%%%%%%
\section{Introduction}
Hydrodynamic turbulence, despite its ubiquitous nature, is extremely complex to be studied analytically. The degree of complexity gets
considerably enhanced when the system consists of a magnetohydrodynamic (MHD) fluid. Yet several theoretical (analytical) works have 
been carried out in the framework of both the abovesaid cases when one assumes the incompressibility (density of the fluid is constant in space
and time and the velocity field is a solenoidal vector) of the fluid. It was Kolmogorov \cite{K41} who for the first time derived in 1941 an exact 
relation for the third-order moment of the velocity structure functions in incompressible hydrodynamic turbulence -- the famous 4/5 law 
\cite[see also][]{eyink}. 
Later two important exact relations were derived respectively for the conduction of a passive scalar in a turbulent system \cite{yaglom} and 
incompressible MHD turbulence \cite{PP98}. Note that exact relations can also be derived for quasi-geostrophic flows \citep{lindborg} 
or dispersive MHD \cite{galtier08,meyrand}, whereas attempts are currently made to model axisymmetric turbulence \cite{galtier12}. 

In a recent Letter \cite{banerjee11} we have derived, for the first time, an exact relation for compressible hydrodynamic turbulence with an isothermal 
analytic closure $P= c^2_s \rho$, where P, $ \rho $ and $c_s$ are respectively the fluid pressure, density and sound speed (for our case it is 
supposed to be constant) defined at every point of the flow field. In the present article we generalize our previous work to the MHD case. 
Of course, the taking into consideration of compressibility of an MHD fluid leads us to get closer to the reality taking place for example in the solar 
wind because (i) it is a plasma (so contains charged species), (ii) Ulysses data analysis has shown clear evidences of the effect of compressibility 
\cite{carbone2009} and (iii) Voyager data analysis at several Astronomical Units (AU) reveals the presence of both fluctuations and jumps with 
spectra steeper than Kolmogorov \cite{burlaga}. 
Moreover turbulence inside the interstellar clouds, being highly compressible (supersonic), also demands a theoretical background in order to 
be properly understood \cite{SC,statobs,filament}. A thorough theoretical work in compressible MHD turbulence is essential (suggested in 
\cite{bertoglio2001}) above all because it will help to analyse and understand the results of Direct Numerical Simulations (DNS) done for compressible 
turbulence \cite{passot,porter1,kritsuk,federrath,schmidt}. 

A number of numerical works have been carried out for the last two decades in the case of compressible hydrodynamic turbulence. 
Being initiated mainly around the years 1990 \cite{passot87,porter2}, the numerical simulations get considerably sophisticated in the last ten years. 
Recently, Kritsuk et al. \cite{kritsuk} have studied three-dimensional (3D) isothermal compressible hydrodynamic turbulence using Piecewise 
Parabolic Method (PPM) for upto $ 2048^3 $ grid points where the r.m.s. Mach number is around $6$. 
%and a solenoidal forcing is used. 
Their study consists of the Helmholtz decomposition of velocity and they obtain a velocity power spectrum around $-2$ ($-1.95$, $-1.92$, $-2.02$ 
respectively for total velocity, solenoidal and 
compressible velocity components) using $1024^3$ grid points whereas they have a $-5/3$ spectrum for the density-weighted fluid velocity 
$\rho^{1/3} v$. Another important numerical work has been accomplished by Federrath et al. \cite{federrath} for an identical (to the previous one) 
system of turbulence using upto $1024^3$ grid points by applying compressible or solenoidal forcing. For two different types of forcing, they have 
obtained remarkably different power spectra (which questions the universality) corresponding to various quantities like the total velocity, its longitudinal 
and transverse components, $\ln \rho$, $\rho^{1/2} v$ or $\rho^{1/3} v$. Total velocity gives a $-1.86$ and a $-1.94$ power spectra in the supersonic 
range for solenoidal and compressible forcings respectively which are close to the Burgers spectrum (in $k^{-2}$). The compensated spectra in the 
subsonic region cannot be affirmed to date to be free from bottleneck effect and so it is hard to conclude anything for that part. However, a $-5/3$ spectra 
is obtained for $\rho^{1/3} v$ even in supersonic range using solenoidal forcing whereas a compressible forcing gives a $-2.1$ spectrum. Here too a 
plausible presence of bottleneck effect prevents us from inferring for subsonic region. 
Recently, Aluie et al. \cite{aluie2012} have provided empirical evidences of kinetic energy cascade (which is not an invariant in compressible turbulence) 
using subsonic and transonic simulations. Such cascading takes place far in the inertial zone due to very fast decay of the pressure-dilatation term which 
dominates at large scales. They do not give, however, any intuition for supersonic turbulence for which {\it a priori} the kinetic energy cascade may not 
be the dominant process. In the meantime, a theoretical work based on coarse-graining \cite{aluienew} shows that for the kinetic energy a range of 
scales independent of the dissipation and the forcing may exist in compressible turbulence. 

3D DNS of compressible MHD turbulence is more difficult to perform since it requires the computation of a heavier system in which several new regimes 
are possible. In this framework, for example Cho and Lazarian \cite{cho2003} implemented a third-order accurate hybrid and essentially non oscillatory 
numerical scheme to solve the isothermal, ideal MHD equations in a periodic box and for obtaining scaling relations they drove turbulence with a 
solenoidal forcing in Fourier space using $256^3$ grid points. In their study, they executed a mode separation method in order to study the individual 
scaling behaviour of each of the present linear modes in a compressible MHD system \ie Alfv\'en mode, slow and fast magnetosonic (MS) modes. 
They reported a $-5/3$ velocity fluctuation spectrum for each of the Alfv\'en and slow MS modes whereas a $-3/2$ spectrum for fast MS mode. The nature 
of spectra obtained is found to depend very weakly on the value of the plasma $\beta$ parameter (the ratio of kinetic pressure to magnetic pressure). 
An approximately $\ell^{2/3}$ ($\ell$ is the length scale) scaling is also obtained for the second-order structure function. 
In \cite{kowal2007} a Kolmogorov spectrum ($-5/3$) was reported for both the velocity ($v$) and density-weighted velocity ($\rho^{1/3} v$) fluctuations for 
subsonic MHD turbulence whereas only the second quantity follows a $-5/3$ spectrum for supersonic turbulence. 

A relevant work to our present paper is that of Kritsuk et al. \cite{kritsuk09} where the authors studied numerically supersonic turbulence in magnetized 
magnetic clouds. The work is realised with grid resolutions from $256^3$ upto $1024^3$ using PPM, an isothermal closure and a solenoidal forcing to 
drive the fluid. Interestingly, they have observed the scale invariance of the total (kinetic $+$ compressible) energy transfer rate in the inertial zone; 
we shall use this concept later in this article for constructing the different spectra. The most interesting and important aspect of their article is to obtain 
numerically the linear scaling in the 4/3-law of incompressible MHD turbulence to the weakly magnetized supersonic turbulence 
-- the former Els\"asser variables $(\zz^\pm)$ being replaced by the density-weighted Els\"asser variables $( \rho^{1/3} \zz^\pm )$. The present article 
will give, by the help of an exact relation for compressible isothermal MHD turbulence, a theoretical support for this generalization. 

The rest of the paper is organized as follows. Section \ref{sec2} introduces the compressible MHD equations and the correlation functions; 
the derivation of the exact relation which is the main result of this paper is given in Section \ref{sec3}; Section \ref{sec4} is devoted to a discussion 
about the main result where in particular we look at the effect of a uniform magnetic field; finally a conclusion is drawn in Section \ref{sec5}.

%%%%%%%%%%%%%%%
\section{Compressible MHD equations}
\label{sec2}

For constituting an exact relation of compressible MHD turbulence of an isothermal plasma, we rewrite the basic equations (except the continuity equation which is kept unchanged) of ideal MHD  in an unusual form \cite{marsch} as follows:
\ba
\p_t \rho + \nabla \cdot ( \rho \vv ) & = & 0 \, , \label{e1} \\
\p_t \vv  + \vv \cdot {\nabla} \, \vv  &=&  \vA \cdot {\nabla} \, \vA  - {1 \over {\rho}} \nabla \left( P +  {1 \over 2} {\rho \vA^2} \right) \nonumber \\
&&-  \vA (\nabla \cdot \vA) + \D + \fz \, , \label{e2} \\
\p_t \vA + \vv \cdot \nabla \, \vA &=& \vA \cdot \nabla \, \vv - {\vA \over 2} (\nabla \cdot \vv)  + \d_M \, , \, \label{e3} \\
 \vA \cdot \nabla \, \rho &=& - 2\rho ( \nabla \cdot \vA )  \, , \label{e4}
\ea
with $\rho$ the plasma density, $\vv$ the fluid velocity field, $\vA=\bb / \sqrt{\mu_0 \rho}$ the Alfv\'en velocity and $P$ the pressure. 
The terms $\D$, $\d_M $ and $\fz$ represent respectively the contribution of the kinetic viscosity, magnetic resistivity and the external forcing. 
It can be identified that equations (\ref{e2}), (\ref{e3}) and (\ref{e4}) represent the force equation, the Faraday's equation (with idealised Ohm's law injected) 
and the zero divergence of magnetic field respectively. For the sake of simplicity in the present paper we use the isothermal closure (\ie $P=C_s^2 \rho$). 
It is important to note that $\vA$ as it is defined in our case does not correspond to the usual Alfv\'en velocity which does not depend on the density but 
just on the mean density. Moreover, the Alfv\'en mode is defined only when there is a mean external magnetic field $\Bz$ whereas our $\vA$  is defined 
independently of the presence of $\Bz$.

As in the hydrodynamic case, our analysis will be carried out in the physical space in terms of two-point correlation functions and structure functions, 
where the unprimed quantities represent the properties at the point $\xx$ and the primed quantities correspond to the point $\xx'$ (with $\xx' =\xx+\rr$). 
Our system is supposed to be statistically homogeneous and to undergo a completely developped turbulence. The analysis is general 
and contains no hypothesis of isotropy. 

In order to establish an exact relation, we shall utilise the conservation of the total energy whose density is given as:
\be
E ( \xx ) =  { \rho \over 2 } ( \vv \cdot \vv +  \vA \cdot \vA ) + \rho e \, ,
\ee
where $e$ accounts for the compressible energy term due to the isothermal closure and is expressed as $C_s^2 \ln ( { \rho / \rho_0} )$, where 
$\rho_0$ is the initial equilibrium density. We however introduce another quantity which we call compressible cross-helicity density and is written as: 
\be
H ( \xx )  = \rho  \vv \cdot \vA \, .
\label{eheli}
\ee
Although this quantity is not conserved in compressible MHD, we need it (as we shall see later) to express the final flux term in a more indicative way 
(in connection with the incompressible exact relations \cite{PP98}). We can write the expressions for $E' ( \xx' ) $ and $H' ( \xx' )$ in an identical way 
as that of the above.

%%%%%%%%%%%%%%%
\section{Derivation of an exact relation}
\label{sec3}

Following the formalism of our previous paper \cite{banerjee11}, we define the relevant two-point correlation functions associated to the total energy 
and the compressible cross-helicity density as:
\ba
R_E &=& { \rho \over 2 } ( \vv \cdot \vv' + \vA \cdot \vA ' ) + \rho e'  , \\
R_H &=&  {\rho \over 2} (\vv \cdot \vA'   +  \vA  \cdot  \vv' ) \, ,
\ea
and similarly for the primed quantities. 
Using the above definitions and introducing compressible Els\"asser variables $\zz^\pm \equiv \vv \pm \vA$, we obtain:
\begin{eqnarray}
({ E + E' })  \pm  ({ H + H' })  &-& ( {R_E \pm  R_H + R'_E  \pm  R'_H} ) \nonumber \\ 
- \delta \rho \delta e &=& {1 \over 2} \delta ( \rho \zz^\pm ) \cdot \delta \zz^\pm \, , 
\end{eqnarray} 
where for any variable $\psi$, $\delta \psi \equiv \psi({\bf x} + \rr) - \psi ({\bf x}) \equiv \psi'-\psi$. 
Under ensemble average (which is equivalent to spatial average for a statistically homogeneous system) the above relations get reduced to:
\ba
{\langle E \rangle \pm  \langle H \rangle}  &-&  {1 \over 2} { \langle {R_E \pm  R_H + R'_E  \pm  R'_H}    \rangle } 
-  {1 \over 2}  \langle \delta \rho \delta e \rangle  \nonumber \\
&=& {1 \over 4 }  \langle  \delta ( \rho \zz^\pm ) \cdot \delta \zz^\pm \rangle \, .
\ea
Now we shall calculate:
\be  
 \p_t    \langle {R_E \pm  R_H + R'_E  \pm  R'_H} \rangle \, ,
\ee 
in order to see how the incompressible relations \cite{PP98} can be generalized to the compressible pseudo-energies 
($e^\pm = \rho \bf z^\pm \cdot \bf z^\pm$) which are no more conserved. 
By using the equations (\ref{e1})--(\ref{e4}) and basic identities of vector calculus, we obtain after some calculations:
\begin{widetext} 
\ba
\p_t   {( \rho \vv \cdot \vv' )} = &-& \nabla \cdot [\rho ( \vv \cdot \vv' ) \vv ]  +  \nabla \cdot [\rho \vA ( \vv' \cdot \vA )] - ( \vv' \cdot \vA ) 
 \left[ \nabla \cdot (\rho \vA ) \right]  - \nabla \cdot (P \vv') -  \nabla \cdot ({1 \over 2} \rho \vA ^2 \vv') \\  
 &-& \rho (\vv' \cdot \vA ) ( \nabla \cdot \vA ) - \nabla' \cdot [\rho (\vv . \vv') \vv']  + \rho (\vv \cdot \vv') (\nabla' \cdot \vv' ) +  
 \nabla' \cdot [\rho (\vv \cdot \vA ' ) \vA ' ]  -  \rho ( \vv \cdot \vA ' ) ( \nabla'  \cdot  \vA' )  \nonumber \\
 &-&  {\rho \over \rho'} {\nabla' \cdot (P' \vv) }  - {\rho \over \rho'} \nabla' ( {1 \over 2} \rho' \vA'^2 \vv ) -  \rho ( \vv \cdot \vA' ) (\nabla' \cdot \vA' ) 
 + d_1 + f_1 \, , \nonumber  \\
\p_t ( \rho \vA \cdot \vA') = &-& \nabla \cdot [\rho (\vA' \cdot \vA ) \vv ]  +  \nabla \cdot [\rho \vA ( \vv \cdot \vA' )]  -  ( \vA' \cdot \vv) [\nabla \cdot (\rho \vA )]  
 - {1 \over 2} \rho (\vA \cdot \vA' ) ( \nabla \cdot \vv ) \\ 
 &-&  {1 \over 2}  \rho (\vA \cdot \vA') ( \nabla' \cdot \vv' ) -  \nabla \cdot [\rho (\vA \cdot \vA') \vv'] + \rho (\vA \cdot \vA')( \nabla' \cdot \vv' ) + 
 \nabla' \cdot [\rho (\vA \cdot \vv') \vA']  \nonumber \\
  &-&  \rho (\vA \cdot \vv' )( \nabla' \cdot \vA' ) + d_2 \, , \nonumber \\
 \p_t (\rho \vv \cdot \vA' ) = &-& \nabla \cdot [\rho \vv ( \vv \cdot \vA' ) ] +  \nabla \cdot [\rho \vA ( \vA \cdot \vA') ] - ( \vA' \cdot \vA )[\nabla \cdot (\rho \vA ) ] -  
 \nabla \cdot ( P \vA' )  -  \nabla \cdot ( {1 \over 2} \rho v_A^2 \vA') \\
 &-&  \rho (  \vA ' \cdot \vA ) (\nabla \cdot \vA ) -  \nabla \cdot [\rho (\vv \cdot \vA ' ) \vv']  + \rho ( \vv \cdot \vA ' ) ( \nabla' \cdot \vv') +  
 \nabla' \cdot [\rho ( \vv \cdot  \vv' ) \vA ' ] \nonumber \\
 &-&  \rho ( \vv \cdot \vv' ) ( \nabla' \cdot \vA ' )  -  {1 \over 2}  \rho ( \vv \cdot \vA ' )( \nabla' \cdot \vv' )   + d_3 + f_2 \, , \nonumber \\
  \p_t (\rho \vA \cdot \vv' ) = &-& \nabla \cdot [\rho \vv ( \vv' \cdot \vA ) ] + \nabla \cdot [\rho  \vA ( \vv' \cdot \vv ) ]  -  ( \vv \cdot \vv' ) [\nabla \cdot (\rho \vA )] 
  - {1 \over 2} \rho (\vv' \cdot \vA )( \nabla \cdot \vv ) \\
  &-&  \nabla' \cdot [\rho ( \vA \cdot \vv' ) \vv' ] + (\nabla' \cdot \vv') (\rho \vA \cdot \vv') + \nabla' \cdot [\rho (\vA \cdot \vA')\vA' ]  -  
  \rho (\vA \cdot \vA') ( \nabla ' \cdot \vA' )  \nonumber \\
  &-&  {\rho \over \rho'}  \nabla' \cdot [P' \vA ]  -  {\rho \over \rho'} \nabla' \cdot [{1 \over 2} \rho'  v_A'^2 \vA ]  -  \rho (\vA \cdot \vA') (\nabla' \cdot  \vA' )  
  + d_4 + f_3 \, , \nonumber \\ 
 \p_t (\rho e') = &-& \nabla' \cdot ( \rho e' \vv' ) -  \nabla \cdot ( \rho e' \vv) - \nabla' \cdot (P \vv') + \rho e' ( \nabla' \cdot \vv' ) \, .\\
\nonumber
\ea
\end{widetext}
where $d_i$ and $f_i$ ($i=1$ to $4$) represent the dissipative and forcing contributions respectively. 
Hence we have all the necessary elements for the dynamic equation of  $R_E$ and $R_H$. Similarly we can write (by symmetry) the expressions 
of $R'_E$ and $R'_H$ just by carefully interchanging the primed and the unprimed quantites. Now if we apply statistical averaging operator on each 
term, after judicious re-arrangement of all the terms we finally obtain:
\begin{widetext} 
\ba
\p_t \langle  R_E \pm R_H &+& R'_E \pm R'_H \rangle = 
   \nabla_r  \cdot \left\langle \left[ {1 \over 2}  \delta ( \rho \zz^\pm) \delta \zz^\pm   +  \delta \rho \delta e  \right] {\delta \zz^\mp}
  + {\overline{\delta} (e  +  {v_A ^2 \over 2}) \delta ( \rho \zz^\pm )} \right\rangle \\
&-&  {1 \over 4} \left\langle  {v_A'^2 \over c_s^2} \nabla' \cdot ( \rho \zz^\pm e') 
  + {v_A^2 \over c_s^2} \nabla \cdot ( \rho' \zz'^\pm e) \right\rangle \nonumber \\
&+&  \left\langle (\nabla \cdot \vv) \left[  R_E' \pm  R_H'  -  E'  \mp  H'  \mp  {{\overline{\delta} \rho}  \over 2} (\vA \cdot \zz'^\pm) 
  - {P'\over 2}  +  {P_M' \over 2} \right] \right\rangle  \nonumber \\
&+& \left\langle (\nabla' \cdot \vv')   \left[  R_E \pm  R_H  -  E  \mp  H  \mp  {{\overline{\delta} \rho}  \over 2} (\vA ' \cdot \zz^\pm)  
  -  {P\over 2}  +  {P_M \over 2}    \right]   \right\rangle  \nonumber  \\
  &+& \left\langle ( \nabla \cdot \vA ) \left[ R_H  \pm  R_E   -  R_H' \mp  R_E'  \pm  E'  +  H'  
   - \overline{\delta} \rho (\vA \cdot \zz'^\pm) \pm  {5P'\over 2}  \pm  {P_M' \over 2}  \right]  \right\rangle \nonumber \\
  &+& \left\langle ( \nabla' \cdot \vA' ) \left[ R_H'  \pm R_E'  - R_H  \mp  R_E \pm  E  +  H  
   - \overline{\delta} \rho (\vA' \cdot \zz^\pm) \pm  {5P\over 2}  \pm  {P_M \over 2}  \right]  \right\rangle + D_\pm + F_\pm  \,  , \nonumber
\ea
\end{widetext}
where $\overline{\delta} X \equiv (X+X')/2$ and $P_M =  (1/2) \rho v_A^2$ is the magnetic pressure. $D_\pm$ and $F_\pm$ represent respectively the 
averaged dissipative and forcing terms. The above expression consists of two equations which represent the generalized form (to compressible 
MHD) of the coupled equations (4) and (6) of \cite{PP98}. As it was already shown \cite{marsch}, for compressible flows the cross-helicity is no longer 
conserved and we cannot (by equation (\ref{eheli})) let the term $\p_t    \langle {R_E \pm  R_H + R'_E  \pm  R'_H} \rangle $ vanish to the extent of a 
stationary state where the average forcing term and the average resultant dissipative (kinetic + magnetic) terms cancel each other to ensure the 
conservation of total energy. Hence for the compressible case, these two equations individually do not lead to an exact relation comprising of a 
universal quantity. However, if we add both we find:
\begin{widetext}
\ba
\p_t \langle R_E + R'_E \rangle &=&  
{1 \over 2}  \nabla_r  \cdot  \left\langle \left[  {1 \over 2}  \delta ( \rho \zz^-) \cdot \delta \zz^-   +  \delta \rho \delta e  \right]  {\delta  \zz^+}
+  \left[  {1 \over 2}  \delta ( \rho  \zz^+) \cdot \delta \zz^+   +  \delta \rho \delta e  \right]   {\delta \zz^-}
+  {\overline{\delta} (e + { v_A^2  \over 2}) \delta ( \rho \zz^- + \rho \zz^+)} \right\rangle \label{toto1}\\
&-& {1 \over 8} \left\langle { v_A'^2 \over c_s^2} \nabla' \cdot ( \rho \zz^+ e')  +  {v_A^2 \over c_s^2} \nabla \cdot ( \rho' \zz'^+ e) 
+ { v_A'^2 \over c_s^2} \nabla' \cdot ( \rho \zz^- e')  +  {v_A^2 \over c_s^2} \nabla \cdot ( \rho' \zz'^- e) \right\rangle \nonumber \\
&+& \left\langle (\nabla \cdot \vv)  \left[ R_E'  -  E'  - { {\overline{\delta} \rho} \over 2} (\vA ' \cdot \vA )  + {P_M' - P' \over 2}  \right] \right\rangle 
+ \left\langle (\nabla' \cdot \vv' )  \left[ R_E -  E  - { {\overline{\delta} \rho} \over 2} (\vA  \cdot \vA ' ) +  {P_M -P \over 2}   \right] \right\rangle 
\nonumber \\
&+& \left\langle ( \nabla \cdot \vA ) \left[ R_H  -  R_H'  +  H' -  \overline{\delta} \rho  ( \vv' \cdot \vA ) \right] \right\rangle 
+ \left\langle ( \nabla' \cdot \vA ' ) \left[ R_H'  -  R_H  +  H   -  \overline{\delta} \rho  ( \vv \cdot \vA' ) \right] \right\rangle + D + F \, , \nonumber
%\nonumber 
\ea
\end{widetext}
where $D = {(D_+ + D_-) /2}$ and $F = {(F_+ + F_-) / 2}$. Now to establish the final relation, we assume that in the limit of infinite (kinetic and magnetic) Reynold's numbers, the system attains a stationary state. This assumption, when applied to the relation (\ref{toto1}), shows that the left hand term of the 
above equation vanishes. Now if we concentrate far in the inertial zone where the dissipative terms are negligible with respect to the other terms, 
we are left with:
\begin{widetext}
\ba
- 2\varepsilon &=&
{1 \over 2} \nabla_r  \cdot  \left\langle \left[  {1 \over 2}  \delta ( \rho \zz^-) \cdot \delta \zz^- +  \delta \rho \delta e  \right]  {\delta  \zz^+} 
+  \left[  {1 \over 2}  \delta ( \rho  \zz^+) \cdot \delta \zz^+   +  \delta \rho \delta e  \right]  {\delta \zz^-} 
+  {\overline{\delta} (e + { v_A^2  \over 2}) \delta ( \rho \zz^- + \rho \zz^+)} \right\rangle \label{toto2} \\
&-&  {1 \over 8} \left\langle { 1 \over \beta' } \nabla' \cdot ( \rho \zz^+ e')  +  {1 \over \beta} \nabla \cdot  ( \rho' \zz'^+ e) 
+  { 1 \over \beta'} \nabla' \cdot ( \rho \zz^- e')  + {1 \over \beta} \nabla \cdot  ( \rho' \zz'^- e) \right\rangle \nonumber \\
&+& \left\langle (\nabla \cdot  \vv)  \left[ R_E'   -  E'  - {{\overline{\delta} \rho} \over 2} (\vA' \cdot  \vA) +  {P_M' -P' \over 2}  \right]  \right\rangle 
+ \left\langle (\nabla' \cdot \vv' )  \left[ R_E   -  E  - {{\overline{\delta} \rho} \over 2} (\vA  \cdot  \vA' ) + {P_M -P \over 2}  \right]  \right\rangle 
\nonumber \\
&+& \left\langle ( \nabla \cdot  \vA ) \left[ R_H  -  R_H'  +  H'   -  \overline{\delta} \rho ( \vv' \cdot  \vA ) \right] \right\rangle 
+ \left\langle ( \nabla' \cdot  \vA' ) \left[ R_H'  -  R_H  +  H   -  \overline{\delta} \rho  ( \vv \cdot  \vA ' ) \right] \right\rangle \, , \nonumber   
\ea
\end{widetext}
where $\varepsilon$ denotes the mean total energy injection rate (which is equal to the mean total energy dissipation rate). In the above equation the 
flux terms are deliberately written in terms of the compressible Els\"asser variables $\zz^\pm$ whereas the source terms are expressed in terms of the 
velocity and the compressible Alfv\'en velocity. This writing is expected to be useful for (i) an attempt to generalize the Alfv\'en effect (introduced by 
Kraichnan) to describe the phenomenology of compressible MHD turbulence and (ii) understanding separately the contribution of the velocity field 
and the Alfv\'en term in the source term. It is interesting to note that in the second part of our flux, the $\beta$ parameter (\ie $c_s^2 / v_A^2$) 
appears which can help us understand (or obtain) the different limits depending upon its value. 
The above equation, being considerably bulky, is laborious to handle. For quick references, we give below a schematic view of that equation:
\be  
-2 \varepsilon = \left\langle \Phi_1 +  {1 \over \beta} \Phi_2 +   (\nabla \cdot \vv) S_1 +  (\nabla \cdot \vA ) S_2 \right\rangle \, \label{toto3}. 
\ee

%%%%%%%%%%%%%%%
\section{Discussion}
\label{sec4}

%%%%%%%%%%%%%%%
\subsection{Relevant limits} 
Equation (\ref{toto2}) is the main result of the paper. It is an exact relation for compressible isothermal MHD turbulence in terms of correlation 
functions. From this equation we can immediately investigate particular limits. 

\subsubsection{Incompressible MHD} 
In this limit, all the source terms vanish. Moreover, the compressible energy term becomes zero (evident from its definition as for 
incompressible case $\rho$ = $\rho_0$) and we get finally (normalising the density to unity):
\be
-2 \varepsilon = {1 \over 4} \nabla_r \cdot \left\langle ( \delta  \zz^-  )^2 {\delta  \zz^+} +  (\delta \zz^+ )^2 {\delta \zz^-}  \right\rangle \, . 
\ee
Note that the term $\nabla_r  \cdot \langle {\overline{\delta}  ({ v_A^2  / 2}) \delta ( \zz^- +  \zz^+)} \rangle $ can be shown to be zero in incompressible case 
by expanding it and letting the $\nabla_r$ enter inside the average operator in a judicious manner. A rewriting of the $\zz^\pm$ in terms of $\vv$ and $\bb$ 
(assuming the constant density to be normalised to unity) help us recognise the primitive form of the equations found in \cite{PP98}.

\subsubsection{Compressible hydrodynamics}  
To verify the hydrodynamic limit, we put $\vA = 0$ and so $ \zz^+ = \zz^- = \vv $; we are left with:
\ba
- 2\varepsilon &=& 
\nabla_r  \cdot  \left\langle \left[  {1 \over 2}  \delta ( \rho \vv) \cdot \delta \vv   +  \delta \rho \delta e  \right]  {\delta  \vv}
+  {\overline{\delta} e \delta ( \rho  \vv)} \right\rangle \nonumber \\
&+& \left\langle (\nabla \cdot \vv)  \left[ R_E'   -  E'   -  {P' \over 2}  \right] \right\rangle \nonumber \\
&+& \left\langle (\nabla' \cdot \vv' )  \left[ R_E   -  E   -  {P \over 2} \right]  \right\rangle \, \label{toto7} . 
\ea
It is nothing but the exact relation which was derived in \cite{banerjee11}. The only difference between the relation obtained above and that in the 
published paper is that here the pressure terms are written as the source terms whereas previously those were considered to contribute in flux terms.

\subsubsection{High and low $\beta $ plasmas} 
Without problem we admit that in the limit where the beta parameter of the plasma tends to infinity (very large value), \ie the plasma becomes 
almost incompressible (although not entirely), the flux term $ \Phi_2 / \beta $ becomes negligible with respect to 
$\Phi_1$  of equation (\ref{toto3}). On the contrary for a very small beta value, where the plasma can be assumed to be cold and magnetised 
(kinetic pressure negligible with respect to magnetic pressure), the term $\Phi_2 / \beta$  dominates over $\Phi_1 $ term and at that situation the 
effective flux term becomes (after rearrangement):
\be
-  {1 \over 4} \left\langle  { 1 \over \beta' } \nabla' \cdot ( \rho \vv e')  +   {1 \over \beta} \nabla \cdot ( \rho' \vv' e)  \right\rangle  \, . 
\ee 

%%%%%%%%%%%%%%%
\subsection{Presence of an external magnetic field} 
Relation (\ref{toto2}) comprises the total magnetic field at each point of the flow field. This total field $\bb$ at each point can be supposed (a 
realistic case) to have a fluctuating part (vary in space and time) $\tilde{\bb}$ superimposed on a constant external magnetic field $\Bz$. 
In the following, we shall investigate the flux and the source terms under the said situation. 

\subsubsection{Flux contribution}
The part of the total flux term which contains $\Bz$ can be expressed as (with $\mu_0 = 1$): 
\begin{widetext}
\ba
\langle \Phi_{\Bz} \rangle &=& 
{\nabla_r \over 2} \cdot \left\langle \delta  \left( {1 \over \sqrt{\rho}} \right)  \delta (\sqrt{\rho}) \left[ \Bz  \times ( \delta \vv \times \Bz ) \right]  
+ \delta ( \sqrt{\rho})  \left[ \Bz \times (\delta \vv \times \delta {\bf \tilde{v}_A} ) \right]  
 - \delta^2 \left( {1 \over \sqrt{\rho}} \right) \left[ \delta (\rho \vv) \cdot \Bz \right] \Bz  \right. \\ 
&+& \left. \delta \left( {1 \over \sqrt{\rho}} \right) \left[ \delta (\rho {\bf \tilde{v}_A} ) \times (\delta \vv \times \Bz ) 
+ \delta (\rho \vv ) \times (\delta {\bf \tilde{v}_A} \times \Bz ) \right] + B_0 ^2 \overline{\delta}  \left( 1 \over  \rho \right) \delta (\rho \vv) 
+  2 \left[ \Bz \cdot \overline{\delta} \left( { {\bf \tilde{v}_A} \over \sqrt{\rho} } \right) \right]  \delta (\rho \vv)  \right\rangle \nonumber \\ 
&-& {1 \over 2}  \left\langle {B_0^2 \over  2 \rho'^2} \rho \vv \cdot \nabla' \rho'  + {{\Bz \cdot \tilde{\bb}} \over  \rho'^2}  \rho \vv \cdot \nabla' \rho' 
+ {B_0^2 \over 2 \rho^2} \rho' \vv' \cdot \nabla \rho + {{\Bz \cdot \tilde{\bb'}} \over \rho^2}  \rho' \vv' \cdot \nabla \rho  \right\rangle  \, ,  \nonumber
\ea
\end{widetext}
where ${\bf \tilde{v}_A}  = \tilde{\bb} / \sqrt{\rho}$. Now we assume the external field $\Bz$ to be very strong so that $B_0 \gg \vert \tilde{\bb} \vert$ 
(and also $B_0 \gg \vert \delta \bf v \vert $).
\ADD{This situation is classical in astrophysics: for example, there are many evidences of magnetic loops in the turbulent solar corona which are 
interpreted as strong uniform magnetic fields on which small magnetic and velocity fluctuations are present \cite{parenti,buchlin}.}
We shall just consider the terms weighted by $B_0^2$. After some straightforward calculations, we obtain 
the resultant flux term (the magnetic terms  without $B_0$ and with single power of $B_0$ are neglected) which writes at main order:
\ba
\langle \Phi_{\Bz} \rangle &=& {\nabla_r \over 2} \cdot \left\langle \delta  
\left( {1 \over \sqrt{\rho}} \right)  \delta (\sqrt{\rho}) \left[ \Bz  \times ( \delta \vv \times \Bz ) \right]  \right. \nonumber \\
&& \left. - \left[\delta \left( {1 \over \sqrt{\rho}} \right)\right]^2 \left[ \delta (\rho \vv) \cdot \Bz \right] \Bz \right\rangle \, . \label{toto4}
\ea
The above expression gives the modifying part of the flux in the presence of a strong constant magnetic field applied externally. 
One can easily understand that the modification is purely due to compressibility. 
(Note that the pure kinetic terms (not shown in (\ref{toto4})) gives also a contribution to the total flux.) 
It is also interesting to 
notice that in expression (\ref{toto4}) the fluctuations are exclusively kinetic in nature (because of the absence of \eg Hall type term in the basic equations). 
One can easily verify that the $\delta \vv$ of the first term and the $ \delta (\rho \vv)$ of the second term of the above expression can be replaced by 
$\delta \vv_\bot$ and $\delta (\rho \vv_\| )$ respectively where $ \delta \vv_\bot   \bot  \Bz $ and $ \delta \vv_\|  \| \Bz $. 
The pure kinetic terms can however be omitted by assuming the external magnetic contribution to be dominant with respect to the velocity and the density fluctuations and then expression (\ref{toto4}) will represent the total flux contribution.

\subsubsection{Source contribution}
The source terms are also modified due to the effect of a strong external magnetic field. At main order (keeping only the terms in $B_0^2$), the terms of 
type $ \langle (\nabla \cdot \vv) S_1 \rangle $ get reduced to:
\be 
\langle \Psi_\vv  \rangle =  { {B_0^2} \over 2} {\left \langle \delta \left( \nabla \cdot \vv \right) \delta 
\left( 1 \over {\sqrt{\rho}} \right) \overline {\delta}(\sqrt{\rho}) - \overline{\delta} ( \nabla \cdot \vv) \right\rangle}  \, , \label{toto5}  
\ee               
and the source terms like $ {\langle (\nabla \cdot \vA ) S_2 \rangle} $ writes:
\be
\langle \Psi_\vA \rangle = 
\ee
$$\Bz \cdot \left \langle  \nabla \left( 1 \over \sqrt{\rho} \right) 
\left[(\Bz \cdot \vv' ) \left\lbrace \rho' \delta {\left( 1 \over \sqrt{\rho} \right)} \right\rbrace 
-  {(\Bz \cdot \vv)  {{\delta \rho} \over {2\sqrt{\rho'} }} }\right] \right.  $$
$$\left. -  \nabla' \left( 1 \over \sqrt{\rho'} \right) \left[(\Bz \cdot \vv ) \left\lbrace \rho \delta {\left( 1 \over \sqrt{\rho} \right)} \right\rbrace   
-  {(\Bz \cdot \vv' )  {{\delta \rho} \over {2\sqrt{\rho} }} }\right] \right\rangle \, . $$
We note that the latter expression implies only parallel components of the velocity.

\subsubsection{Reduced anisotropic law}

Further simplifications are possible if we assume that the velocity field vector at each point of the flow field is (at the main order) perpendicular to the 
external magnetic field \ie if $ v_\| =v'_\| = 0 $, and therefore $ \vv \equiv \vv_\perp $. In that case, $ \langle \Psi_{\vA} \rangle $ vanishes and so is the 
second term of $\langle \Phi_{Bz} \rangle$; then the corresponding reduced exact relation can be simply written as:
\be
- 4 \varepsilon  =   B_0^2 {\nabla_{r_{\perp}}} \cdot \left\langle \delta  
\left( {1 \over \sqrt{\rho}} \right)  \delta (\sqrt{\rho}) \delta \vv_\perp  \right\rangle \label{toto6} 
\ee
$$
- { {B_0^2} \over 2} {\left \langle \left( \nabla_\perp \cdot \vv_\perp \right) \left(1 + \sqrt{\rho \over \rho' }\right) 
+ \left( \nabla'_\perp \cdot \vv'_\perp \right)  \left( 1 + \sqrt{\rho' \over \rho} \right) \right \rangle} \, , 
$$
where $\nabla_\perp$ implies derivatives transverse to $\Bz$. 
Expression (\ref{toto6}) is the second main result of this paper: it is a limiting case of the exact relation (\ref{toto2}) for isothermal compressible MHD 
turbulence under the influence of a strong external uniform magnetic field (sub-alfv\'enic turbulence regime). 
\ADD{We may simplify the previous equation by assuming axisymmetry; the exact relation (\ref{toto6}) can be written symbolically as 
(by using cylindrical coordinates)
\begin{equation}
- 4 \varepsilon = {1 \over r_\perp} \partial_{r_\perp} (r_\perp {\cal F}_{r_\perp}) + {\cal S} \, , \label{THI}
\end{equation}
where ${\cal F}_{r_\perp}$ the radial component of the energy flux vector (first term in the right hand side of (\ref{toto6})) and 
${\cal S}$ is a source/sink term (last term in the right hand side of (\ref{toto6})). If we define an effective mean total energy injection rate as 
$\varepsilon_{\rm{eff}} \equiv \varepsilon+{\cal S} /4$, a simple interpretation of expression (\ref{THI}) can be proposed as we see in 
Fig. \ref{Fig1}: whereas for a direct cascade the energy flux vectors are oriented towards the axis of the cylinder, dilatation and 
compression are additional effects which act respectively in the opposite or in the same direction as the flux vectors (since terms like, 
$1 + \sqrt{\rho' / \rho}$, are positive).}

\subsubsection{Phenomenology and energy spectra} 

In this section we shall make a prediction on compressible spectra using the derived theoretical relation (\ref{toto6}): although simplified, it is 
considerably indicative for phenomenological intuition. 

The existence of strong external magnetic field $ \Bz $ renders the energy transfer in the parallel (along $ \Bz $) direction negligible in comparison to 
the transverse one \cite{matt96,galtier00,alex,galtier09,matt09,bigot11}. 
Then, our prediction will be made for transverse spectra (\ie with a spectral dependence only in $k_\perp$). From the flux term of expression (\ref{toto6}), 
we find dimensionally (keeping the source terms aside):
\be
\varepsilon \sim  {\rho_\ell v_{A \ell}^2 v_{\perp \ell} \over \ell_\perp}  \sim {\rho_\ell v_{A \ell}^2 \over \ell_\perp / v_{\perp \ell}} \, .
\ee
The above expression corresponds to a compressible phenomenology in which the magnetic energy density, $\rho_\ell v_{A \ell}^2$, is transferred through 
the scales at the transfer time, $\tau_\ell \sim \ell_\perp / v_{\perp \ell}$, and the transfer rate $\varepsilon$. 
We may define the weighted variables: 
\ba
{\cal U}_\ell  &\equiv& {\rho_\ell}^{1/3} v_{\perp \ell}  \, ,  \\
{\cal B}_\ell &\equiv& {\rho_\ell}^{1/3} v_{A \ell} = {\rho_\ell}^{-1/6} B_0 \, ,
\ea
which leads to:
\begin{equation}
\varepsilon \sim {\cal B}_\ell^2 {\cal U}_\ell k_\perp \sim E^{{\cal B}}(k_\perp) k_\perp \sqrt{E^{{\cal U}}(k_\perp) k_\perp} \, k_\perp \, . 
\end{equation}
Whence, the spectral prediction:
\begin{equation}
E^{\cal B} (k_\perp) \sqrt{E^{\cal U}(k_\perp)} \sim \varepsilon k_\perp^{-5/2} \, . 
\end{equation}

\begin{figure}
\resizebox{85mm}{!}{\includegraphics{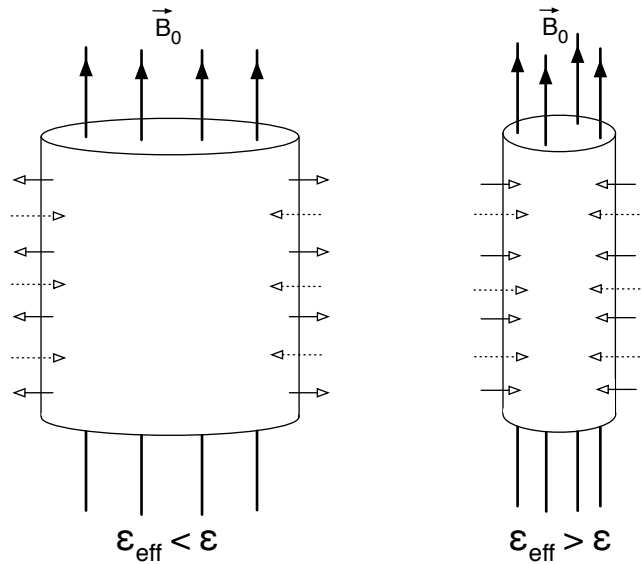}}
\caption{Dilatation (left) and compression (right) phases in space correlation for axisymmetric MHD turbulence. In a direct 
cascade scenario the flux vectors (dotted arrows) are oriented towards the axis of the cylinder. Dilatation and compression (solid arrows) 
are additional effects which act respectively in the opposite or in the same direction as the flux vectors. 
\label{Fig1}}
\end{figure}

Taking into consideration the source terms will only modify $\varepsilon$ for giving an effective mean total energy injection rate 
$\varepsilon_{\rm{eff}}$ (likewise in compressible hydrodynamics \cite{banerjee11}). A compressional effect ($\nabla \cdot \bf v < 0$) will 
increase the effective energy flux rate whereas a dilatational effect ($ \nabla \cdot \bf v >0$) will reduce $\varepsilon_{\rm{eff}}$ as it is 
evident from the source terms of expression (\ref{toto6}). 
\ADD{As stated before a simple picture emerges (see Fig. \ref{Fig1}) with the} additional hypothesis of axisymmetry (symmetry in the plane perpendicular to $\Bz$). 
The flux vectors (dotted arrows) are always oriented towards the axis of the cylinder since a direct cascade is expected. Dilatation and compression 
act additionally (solid arrows): in the first case, the effect is similar to a decrease of the local mean total energy transfer rate whereas in the 
second case it is similar to an increase of the local mean total energy transfer rate. Then, the final prediction for compressible turbulence under a strong
uniform magnetic field is: 
\be
E^{\cal B} (k_\perp) \sqrt{E^{\cal U}(k_\perp)} \sim \varepsilon_{\rm{eff}} (k_\perp) k_\perp^{-5/2} \, ,
\ee 
with {\it a priori} a possible dependence of the effective mean total energy transfer rate on $k_\perp$. It that case, a power law steeper than $-5/2$ 
may be observed at large scales when compressible MHD turbulence becomes supersonic. 

%%%%%%%%%%%%%%%
\section{Conclusion} 
\label{sec5}

The present paper is a generalization of a recent work made for compressible hydrodynamics \cite{banerjee11} in which it was possible to give a 
simple interpretation of the different terms of the exact relation. The introduction of both the Lorentz force and the induction equation renders the 
situation more difficult to analyze but it is believed that the main physical characteristic of compressible MHD emerges when a uniform magnetic field 
is included. In that case, we have derived rigorously a reduced expression (\ref{toto6}) which can be used easily either to analyze direct numerical 
simulations or {\it in situ} observations like those obtained in the solar wind. Since most of the natural plasmas are in interaction with a large scale 
magnetic field, it is likely that the reduced form is the most interesting one. 

Our work confirms the relevance of density-weighted Els\"asser variables, $\rho^{1/3} \zz^\pm$, for compressible MHD turbulence. 
However, unlike the hydrodynamic case, here one should not replace $\zz^\pm$ by $\rho^{1/3} \zz^\pm$ for recovering the incompressible 
scaling laws. Instead the appropriate variable should be, $\rho^{1/3} ({\zz^+}^2 \zz^-  + {\zz^-}^2 \zz^+ )^{1/3}$, for obtaining a satisfactory energy 
scaling as suggested by relation (\ref{toto2}). 
The said relation holds good for any type of compressible MHD turbulence (subsonic, transonic, supersonic, subAlfv\'enic or superAlfv\'enic). 
Hence it can directly be used to analyze direct numerical simulations in compressible MHD turbulence at different Mach numbers or for plasmas 
with different $\beta$ parameter in order to understand the underlying phenomenology. 
In order to approach the astrophysical turbulence more efficiently we have to take into account the gravitational force and also a polytropic closure 
although they would increase enormously the complexity of the analysis.   

It was mentioned in the past that constant flux solutions are in general not relevant for compressible turbulence because the coupling to a sonic field 
provides a supplementary sink which will modify the inertial index \cite{kadomtsev,moiseev,passot88}. A modification of the solenoidal energy spectrum 
(for hydrodynamics turbulence) was proposed by introducing a simple {\it ad hoc} model in which the main compressible parameter was the Mach number. 
Here, we may reach a similar conclusion with a rigorous analysis in terms of two-point correlation functions. We believe that our interpretation in terms 
of effective energy flux rate, source and flux terms are the key ingredients to further understanding of compressible turbulence.

%%%%%%%%%%%%%%%
\paragraph*{Acknowledgment.}
We acknowledge T. Passot for useful discussions. 

%%%%%%%%%%%%%%%

\end{document}